\documentclass{article}
\usepackage{style,times}
\usepackage{amsmath,amssymb,booktabs,tabularx,array}
\usepackage{enumitem,microtype,graphicx,float}
\usepackage{hyperref,url}
\usepackage{longtable,wrapfig}
\hypersetup{hidelinks}
\newcolumntype{Y}{>{\raggedright\arraybackslash}X}

\title{\raggedright ElectrolyteMD-Bench: How Well Do AI Agents Conduct Molecular Dynamics Studies across Electrolyte Solvation Regimes?}
\preprintfinalcopy
\author{\phantom{Anonymous authors}\\\phantom{Paper under double-blind review}}
\AddToShipoutPictureFG*{\AtPageUpperLeft{%
\put(\LenToUnit{143.7958984375bp},-\LenToUnit{163.0bp}){\makebox[0pt][l]{\fontfamily{ptm}\bfseries\fontsize{11.0bp}{11.0bp}\selectfont Shukai Wu}}%
\put(\LenToUnit{196.6904296875bp},-\LenToUnit{159.0bp}){\makebox[0pt][l]{\fontfamily{ptm}\mdseries\fontsize{7.0bp}{7.0bp}\selectfont 1}}%
\put(\LenToUnit{212.1904296875bp},-\LenToUnit{163.0bp}){\makebox[0pt][l]{\fontfamily{ptm}\bfseries\fontsize{11.0bp}{11.0bp}\selectfont Shuo Niu}}%
\put(\LenToUnit{255.9111328125bp},-\LenToUnit{159.0bp}){\makebox[0pt][l]{\fontfamily{ptm}\mdseries\fontsize{7.0bp}{7.0bp}\selectfont 1}}%
\put(\LenToUnit{271.4111328125bp},-\LenToUnit{163.0bp}){\makebox[0pt][l]{\fontfamily{ptm}\bfseries\fontsize{11.0bp}{11.0bp}\selectfont Zhaoming Xu}}%
\put(\LenToUnit{336.51416015625bp},-\LenToUnit{159.0bp}){\makebox[0pt][l]{\fontfamily{ptm}\mdseries\fontsize{7.0bp}{7.0bp}\selectfont 2}}%
\put(\LenToUnit{352.01416015625bp},-\LenToUnit{163.0bp}){\makebox[0pt][l]{\fontfamily{ptm}\bfseries\fontsize{11.0bp}{11.0bp}\selectfont Yan Luo}}%
\put(\LenToUnit{393.2802734375bp},-\LenToUnit{159.0bp}){\makebox[0pt][l]{\fontfamily{ptm}\mdseries\fontsize{7.0bp}{7.0bp}\selectfont 1}}%
\put(\LenToUnit{408.7802734375bp},-\LenToUnit{163.0bp}){\makebox[0pt][l]{\fontfamily{ptm}\bfseries\fontsize{11.0bp}{11.0bp}\selectfont Wentao Lin}}%
\put(\LenToUnit{464.7041015625bp},-\LenToUnit{159.0bp}){\makebox[0pt][l]{\fontfamily{ptm}\mdseries\fontsize{7.0bp}{7.0bp}\selectfont 1}}%
\put(\LenToUnit{164.4541015625bp},-\LenToUnit{173.0bp}){\makebox[0pt][l]{\fontfamily{ptm}\mdseries\fontsize{7.0bp}{7.0bp}\selectfont 1}}%
\put(\LenToUnit{167.9541015625bp},-\LenToUnit{177.0bp}){\makebox[0pt][l]{\fontfamily{ptm}\mdseries\fontsize{10.0bp}{10.0bp}\selectfont  The Hong Kong University of Science and Technology (Guangzhou)}}%
\put(\LenToUnit{261.373046875bp},-\LenToUnit{187.0bp}){\makebox[0pt][l]{\fontfamily{ptm}\mdseries\fontsize{7.0bp}{7.0bp}\selectfont 2}}%
\put(\LenToUnit{264.873046875bp},-\LenToUnit{191.0bp}){\makebox[0pt][l]{\fontfamily{ptm}\mdseries\fontsize{10.0bp}{10.0bp}\selectfont  University of Macau}}%
}}

\newcommand{\Li}{\ensuremath{\mathrm{Li}^{+}}}
\newcommand{\FSI}{\ensuremath{\mathrm{FSI}^{-}}}

\usepackage[T1]{fontenc}
\hypersetup{pdftitle={ElectrolyteMD-Bench},pdfauthor={Shukai Wu; Shuo Niu; Zhaoming Xu; Yan Luo; Wentao Lin}}
\begin{document}
\maketitle\fancyhead{}\renewcommand{\headrulewidth}{0pt}
\raggedbottom

\begin{abstract}
AI agents are advancing the automation of molecular dynamics (MD) research, with the potential to accelerate discovery and design of lithium-battery electrolytes. However, whether they can autonomously complete scientifically reliable electrolyte MD studies remains insufficiently evaluated. We introduce ElectrolyteMD-Bench, which unifies dilute, high-concentration, and localized high-concentration electrolytes in a comparative study across distinct solvation regimes. The benchmark connects local coordination, ionic association, single-particle dynamics, and collective transport within one continuous research process. Agents autonomously perform system construction, equilibration, production sampling, property analysis, and cross-system interpretation, while deciding whether to correct errors, continue sampling, or stop. We audit each study along four dimensions: simulation validity, analysis validity, evidence integrity, and outcome calibration. Across eight model--harness configurations, all 24 system runs produced valid production trajectories, yet none of the eight studies satisfied all scientific requirements. Failures included missing analyses, incorrect physical definitions or numerical implementations, insufficient statistical support, and completion claims inconsistent with retained evidence. These results expose a gap between successful MD workflow execution and reliable scientific study completion. ElectrolyteMD-Bench therefore provides a foundational test and defines a key capability threshold that autonomous electrolyte simulation must cross to advance toward computation-first materials exploration.

\end{abstract}

\section{Introduction}

High-performance lithium-ion batteries are a key technology for transport electrification and large-scale energy storage, with lithium-ion solvation and transport in electrolytes critically influencing battery performance \citep{chang2026electrolyte}. Molecular dynamics (MD) simulations directly track the atomic-scale motion of ions and solvents, providing an important computational means to connect local solvation structure and ionic association with long-range transport. However, conducting a scientifically reliable electrolyte MD study remains technically demanding. System construction, force-field setup, energy minimization, equilibration, and production simulations are interdependent, while subsequent analyses of structure, dynamics, and transport involve distinct physical definitions, sampling timescales, and statistical requirements \citep{yao2022applying}.

Through language-based reasoning and direct interaction with computational environments, artificial intelligence agents can, in principle, convert scientific objectives into executable computational plans. They can coordinate existing tools and adjust subsequent steps in response to intermediate results or execution failures \citep{anand2026mdarena}. However, automating MD requires more than simply running simulations. Agents must ensure that the research workflow and its conclusions remain scientifically valid. This raises a central question: can agents autonomously complete an electrolyte MD study from end to end?

Liquid-electrolyte MD spans several interconnected levels of analysis, each with distinct physical definitions and statistical requirements. Dilute electrolytes (DE), high-concentration electrolytes (HCE), and localized high-concentration electrolytes (LHCE) constitute a representative set of solvation regimes. As salt concentration increases, the local environment can evolve from solvent-dominated structures characterized by solvent-separated ion pairs toward anion-rich structures dominated by contact ion pairs and aggregates \citep{yamada2019advances}. In LHCE, adding a weakly coordinating diluent changes the overall concentration and rheological properties while potentially preserving the characteristic local solvation environment of HCE \citep{chen2018lhce,weintz2026heat}. Differences among these systems therefore extend beyond composition and density to the first coordination shell, ionic association, higher-order aggregation, and the resulting dynamics and collective transport. Previous studies show that viscosity or single-particle self-diffusion alone cannot explain ion transport in HCE and LHCE \citep{bergstrom2024ion}. Ion--ion cross-correlations, ligand exchange, and the formation and disruption of percolating ionic networks can substantially alter conductivity and transference numbers \citep{mohanakrishnan2026correlated,chang2026electrolyte}. Addressing these levels of analysis requires agents to go beyond generating stable trajectories or reporting selected local structural quantities. They must analyze radial distributions and coordination, ion pairing and aggregation, single-particle diffusion, and collective transport, including ionic cross-correlations, using consistent physical definitions. They must also determine whether the available trajectories provide sufficient statistical support for each property.

Motivated by these requirements, we introduce ElectrolyteMD-Bench to investigate whether artificial intelligence agents can autonomously complete an entire electrolyte MD study (Figure~\ref{fig:overview}). The benchmark defines a unified research task across dilute, high-concentration, and localized high-concentration electrolytes. Agents autonomously perform system construction, energy minimization, equilibration, production sampling, and analyses of local structure, ionic association, single-particle dynamics, and collective transport, while using intermediate results to choose simulation protocols, recover from errors, allocate computational resources, and decide when to stop. To assess whether a study is complete, we introduce the SAEO evaluation protocol, covering simulation validity, analysis validity, evidence integrity, and outcome calibration. The protocol independently audits agents' simulation configurations, logs, trajectories, analysis implementations, numerical results, run records, and final reports.

In ElectrolyteMD-Bench, all evaluated agents generated valid production trajectories, but none of the resulting studies satisfied all scientific requirements. The main shortcomings concerned the validity of analysis methods, statistical support for transport properties, and consistency between final research judgments and the available evidence. Agents could misclassify methodological errors as insufficient sampling and even declare a study complete while explicitly acknowledging that required analyses remained unfinished. These findings indicate that the reliability of autonomous research depends not only on completing computations, but also on obtaining reliable property estimates, maintaining complete and consistent research evidence, and ensuring that subsequent actions and final judgments remain constrained by that evidence: correct methods when necessary, extend sampling when insufficient, or acknowledge when a conclusion is not supported.

\begingroup
\setlength{\intextsep}{5pt}
\begin{figure}[H]
\centering
\includegraphics[width=\linewidth]{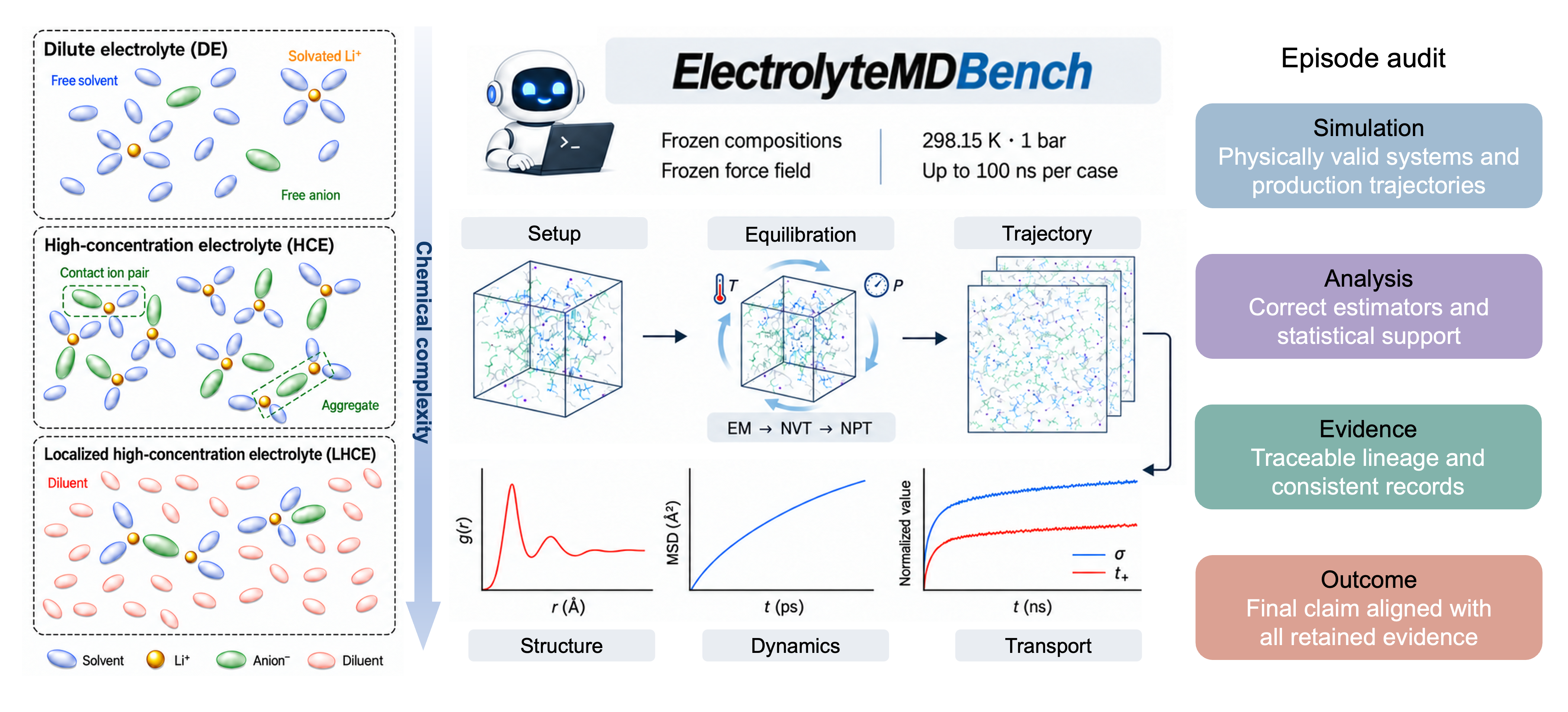}
\caption{Overview of ElectrolyteMD-Bench. }
\label{fig:overview}
\end{figure}
\endgroup

\section{Related Work}

Large language model agents have automated multiple stages of MD workflows. MDCrow integrates domain-tools for protein-system preparation, simulation execution, and trajectory analysis \citep{campbell2026mdcrow}. NAMD-Agent extends automation to solution-phase and membrane-protein systems, covering system construction, NAMD simulations, and standard trajectory analyses (Chandrasekhar \& {Barati Farimani}, 2025). For protein--ligand systems, DynaMate integrates research planning, simulation execution, and analyses within a multi-agent workflow \citep{guilbert2025dynamate}. PolyJarvis extends automated MD to polymers, autonomously constructing and equilibrating systems, calculating thermal and mechanical properties \citep{zhao2026polyjarvis}. MDAgent2 focuses more specifically on MD knowledge and LAMMPS code generation \citep{shi2026mdagent2}. Together, these studies cover several key stages from system construction and simulation execution to property analysis.

Other studies have begun systematically evaluating agents' capabilities in MD tasks. MDGym uses simulation and analysis tasks spanning multiple systems and physical quantities, comparing target properties with reference results. Its findings reveal a substantial gap between executing simulations and obtaining correct results \citep{kumar2026mdgym}. MDArena additionally evaluates the research process through tasks drawn from real biomolecular research projects, emphasizing system preparation, analysis, and post-processing. Its results show that agents may successfully execute individual operations and produce results, yet still fail to complete the overall task because they omit essential steps or lack sufficient verification. \citep{anand2026mdarena}. Still, the coverage of long-timescale stochastic production simulations and complete end-to-end execution remains limited.~

Representative agent-based MD studies have predominantly addressed biomolecular systems, where research commonly concerns conformations, molecular contacts, binding processes, and structural stability. Electrolyte solutions additionally require examining ion--solvent coordination competition, ionic association, and their effects on bulk properties \citep{yamada2019advances,bergstrom2024ion}. Automated MD has recently been applied to electrolyte research. Nakamura et~al. (2026) developed a high-throughput electrolyte MD framework based on expert-defined workflows. Reasoning-to-Simulation further introduces agents into battery-electrolyte simulation and formulation discovery, creating a closed-loop optimization process through reasoning, simulation, and feedback (Ru\v{z}a \& Gómez-Bombarelli, 2026). These studies demonstrate that electrolyte MD workflows can be automated and integrated into agent-driven materials discovery, yet it remains unclear whether agents can reliably conduct a complete electrolyte MD study.

ElectrolyteMD-Bench therefore evaluates a complete electrolyte study rather than a collection of independent, clearly bounded MD tasks. It evaluates agents across three representative electrolyte solvation regimes, covering DE, HCE, and LHCE within a single comparative study. We assess not only the correctness of individual simulations and analyses, but also whether the overall study is scientifically supported by valid methods, sufficient statistical evidence, traceable results, and evidence-consistent conclusions.

\section{ElectrolyteMD-Bench Design}

\subsection{Systems and Agent Configurations}

\begin{wraptable}{r}{0.50\textwidth}
\vspace{-40pt}
\caption{Compositions of MD systems.}
\label{tab:systems}
\centering
\footnotesize
\setlength{\tabcolsep}{2.3pt}
\begin{tabular}{lrrrrr}
\toprule
System & \Li{} & \FSI{} & DMC & TTE & Atoms \\
\midrule
DE & 384 & 384 & 3,840 & 0 & 49,920 \\
HCE & 1,470 & 1,470 & 2,940 & 0 & 49,980 \\
LHCE & 960 & 960 & 1,920 & 960 & 49,920 \\
\bottomrule
\end{tabular}
\vspace{-8pt}
\end{wraptable}

Each complete study episode includes three systems: DE, HCE, and LHCE. Their LiFSI:DMC:TTE molar ratios are 1:10:0, 1:2:0, and 1:2:1, respectively (Table 1) \citep{mohanakrishnan2026correlated}. The target thermodynamic conditions are 298.15 K and 1 bar.~ All simulations use GROMACS \citep{pall2020gromacs} and the same fixed-charge, nonpolarizable, OPLS-AA-compatible molecular model \citep{jorgensen1996opls}. \Li{} and \FSI{} parameters are based on CL\&P \citep{gouveia2017ionic}, DMC uses OPLS-AA-compatible parameters with predefined CHELPG charges, and TTE parameters are assembled from Kim et al. (2023). Charges on \Li{} and \FSI{} are uniformly scaled to 0.72 to approximately account for electronic screening in the nonpolarizable model \citep{leontyev2011polarization}. Parameter acquisition and force-field selection are outside the benchmark scope. All candidates therefore receive identical molecular inputs and task specifications, but no preconstructed systems, simulation protocols, trajectories, analysis scripts, or reference results.

Two coding-agent harnesses are evaluated: Codex, a frontier harness, and OpenCode, an open-source harness. Codex is paired with GPT-5.6-Sol, GPT-5.6-Luna, DeepSeek-V4-Flash, and DeepSeek-V4-Pro, while OpenCode is paired with GPT-5.6-Luna, GLM-5.2, MiniMax-M3, and DeepSeek-V4-Flash. The model set covers flagship and efficiency-oriented variants and includes both proprietary and open-weight models. GPT-5.6-Luna and DeepSeek-V4-Flash are evaluated under both harnesses, enabling matched cross-harness comparisons.

\subsection{MD Task}

Under fixed inputs, agents autonomously conduct system construction, minimization, equilibration, production sampling, analysis, and final interpretation. They determine the simulation protocol and stage-entry and stopping criteria, and must establish system validity and equilibration using quantitative diagnostics rather than predetermined simulation times \citep{merz2018physical,chodera2016equilibration}. Each system has a cumulative budget of 100 ns, including equilibration, production, failed attempts, recovery, validation, and extensions. Within the remaining budget, agents may revise their workflow and extend sampling as needed. No human scientific guidance is provided during formal episodes.

\subsection{Electrolyte Analysis}

\begingroup
\setlength{\intextsep}{5pt}
\begin{table}[H]
\caption{Required analyses and target quantities.}
\label{tab:analyses}
\centering\small
\begin{tabularx}{\linewidth}{@{}p{0.27\linewidth}Y@{}}
\toprule
Analysis category & Target quantities \\
\midrule
Thermodynamics & Density, temperature, pressure where appropriate for the ensemble, potential energy, total energy, and volume \\
Local structure & Radial distribution functions and coordination numbers \\
Ionic association & Contact ion pairs, solvent-separated ion pairs, and aggregates \\
Single-particle dynamics & Self-diffusion coefficients of each species \\
Collective transport & Viscosity, conductivity including ionic cross-correlations, and an explicitly defined transference number \\
\bottomrule
\end{tabularx}
\end{table}
\endgroup

Agents must analyze thermodynamics, local structure, ionic association, single-particle dynamics, and collective transport. For LHCE, agents must separately analyze \Li{} coordination with DMC carbonyl oxygens, \FSI{} oxygens, and TTE ether oxygens. Local environments must be interpreted using coordination numbers, species abundances, and contact or aggregation statistics. Nernst--Einstein conductivity may be reported as an additional result but must be explicitly labeled \citep{francelanord2019correlations}. Collective conductivity used in the main comparisons must include ionic cross-correlations \citep{francelanord2019correlations,fang2023transference}.

Before interpreting results across systems, agents must record the initial exclusion interval, principal atom selections, contact and aggregation definitions, fitting rules, and stopping criteria. After excluding the initial interval, the remaining trajectory is divided into four consecutive, equally long, nonoverlapping time blocks. Both whole-interval estimates and block results must be reported. Blocks assess time dependence, fluctuations, and statistical stability. Their independence must be considered in relation to correlation times \citep{grossfield2018uncertainty}.

Sampling adequacy is assessed separately for each property, rather than using one trajectory length as a universal convergence criterion \citep{wan2021uncertainty}. Diffusion analysis must check whether the fitting interval lies in the diffusive regime. Viscosity analysis must establish a defensible plateau or asymptotic regime for the chosen estimator; for Green--Kubo analysis, this includes examining the stress-correlation integral \citep{maginn2018transport}. Collective-transport analysis must examine the estimates and their stability across blocks. A single trajectory can therefore provide different degrees of statistical support for different properties.

When an estimator is valid but sampling is insufficient, agents must explicitly report this limitation and the corresponding differences between blocks. Such limitations must not be conflated with analytical errors. Items with invalid upstream simulations or stages not yet reached are recorded as upstream-invalid or not reached, respectively. Cross-system comparisons may use only results with consistent definitions and valid supporting data.

\subsection{SAEO Evaluation}

Before agent execution, we established the scientific task requirements and basic evaluation principles: coverage of required observables, simulation and analysis validity, sampling adequacy, evidence traceability, and consistency between final claims and evidence. Because agents autonomously choose simulation and analysis methods, the open-ended task admits implementations and failure modes that cannot be exhaustively specified in advance. We therefore further specified the rubric's operational criteria based on the retained artifacts, defining checks for different implementations, numerical anomalies, and analysis failures. Appendix A.2.4 maps the original requirements to these refinements and their scope.

An LLM judge examines each configuration's archived artifacts item by item and produces preliminary judgments with evidence references \citep{zheng2023judge,liu2023geval,kim2024prometheus}. Domain experts review these judgments and their supporting evidence, focusing on physical definitions, numerical implementations, sampling adequacy, and disputed items, and confirm the final adjudication. The evidence includes original configurations, simulation logs, final trajectories, analysis implementations, run records, and final reports. Evaluation proceeds along four dimensions, which jointly determine the complete-study status.

\begingroup
\setlength{\intextsep}{5pt}
\begin{table}[H]
\caption{The four dimensions of SAEO evaluation.}
\label{tab:saeo}
\centering\small
\begin{tabularx}{\linewidth}{@{}p{0.22\linewidth}Y@{}}
\toprule
Dimension & Verification scope \\
\midrule
Simulation validity & Composition and model consistency, energy minimization, equilibration diagnostics, and validity of production trajectories \\
Analysis validity & Target-quantity coverage, physical definitions, formulas and numerical implementations, trajectory processing, statistical adequacy, and uncertainty \\
Evidence integrity & Correspondence and consistency among inputs, simulation branches, analysis versions, numerical results, and final reports \\
Outcome calibration & Identification and classification of unresolved problems, subsequent actions or stopping decisions, and agreement between final completion claims and available evidence \\
\bottomrule
\end{tabularx}
\end{table}
\endgroup

Incomplete analyses are further separated into three categories: (1) missing required estimators or results; (2) deterministic physical or numerical errors; and (3) valid estimators with insufficient sampling. Final completion claims are evaluated separately for consistency with the available evidence and study status.

The dimensions are recorded separately, with individual items rated as satisfied, partially satisfied, or not satisfied. Partial satisfaction is assigned when an item is only partly supported by the available analysis or evidence, or when a valid estimator is limited by insufficient sampling. Missing required estimators or results, deterministic methodological or numerical errors, and unavailable evidence are rated as not satisfied. The composite score equally weights the four dimension means:
\begin{equation}
\mathrm{Score}
= \frac{100}{4}\sum_{d\in\{S,A,E,O\}}
\left(\frac{1}{n_d}\sum_{i=1}^{n_d}s_{di}\right).
\label{eq:saeo-score}
\end{equation}
Here, $s_{di}\in\{1,1/2,0\}$ denotes satisfied, partially satisfied, or not satisfied, respectively, and $(n_S,n_A,n_E,n_O)=(4,7,5,5)$ gives the number of items in each dimension.

\section{Results and Analysis}

\subsection{The Gap between Valid Production Trajectories and Complete Studies}

Across the evaluated model--harness combinations, all configurations satisfied the criteria for the Simulation dimension (Figure 2). Codex + GPT-5.6-Sol achieved the highest overall score, with its advantage mainly arising from more balanced performance across the subsequent Analysis, Evidence, and Outcome stages. However, none of the eight configurations, including this top-performing combination, ultimately produced an MD study that satisfied all scientific requirements. Current AI agents can already reliably perform technical operations in an MD workflow, but a substantial gap remains before they can complete a scientifically valid MD study. Generating valid MD production trajectories is insufficient, with the main challenges lying in valid analysis, adequate statistical support, traceable and internally consistent evidence, and evidence-consistent final judgments.

\begingroup
\setlength{\intextsep}{5pt}
\begin{figure}[H]
\centering
\includegraphics[width=\linewidth,height=0.73\textheight,keepaspectratio]{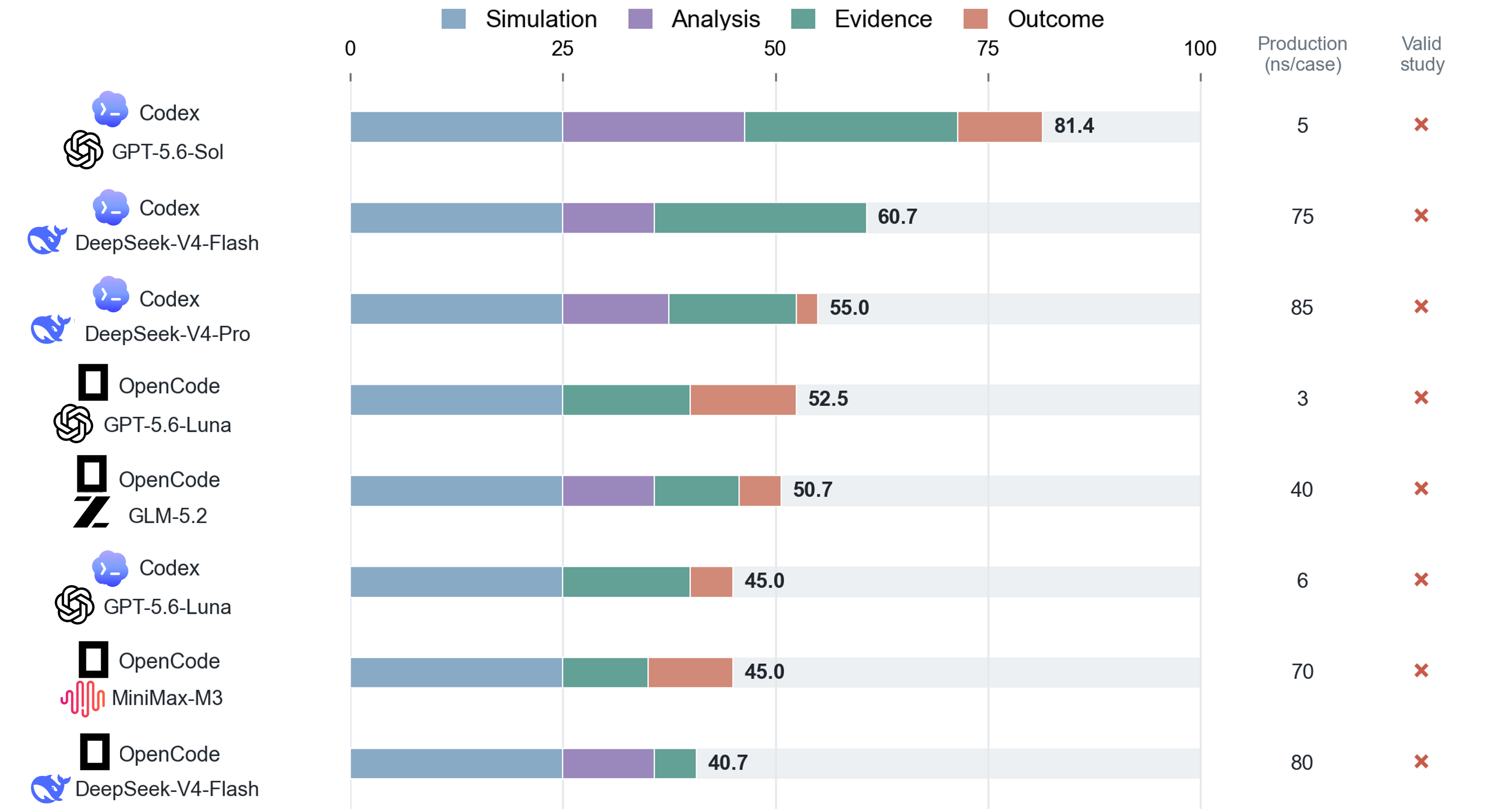}
\caption{Overall scores for eight configurations.}
\label{fig:overall}
\end{figure}
\endgroup

Performance varied substantially across model--harness combinations. Codex + GPT-5.6-Sol, combining a frontier harness with its flagship model, came closest to completing the full study. Beyond this combination, performance did not follow a simple ordering between flagship and efficiency-oriented model variants. This pattern suggests that higher nominal reasoning capability does not consistently translate into better end-to-end scientific performance, and that different stages of the workflow may benefit from different model profiles. It motivates future investigation of stage-adaptive model selection, in which higher-capability and efficiency-oriented models are selected according to the demands of individual research stages.

\subsection{Four-Dimensional Criteria Distinguish Different Failure Types}

\begin{figure}[H]
\centering
\includegraphics[width=\linewidth]{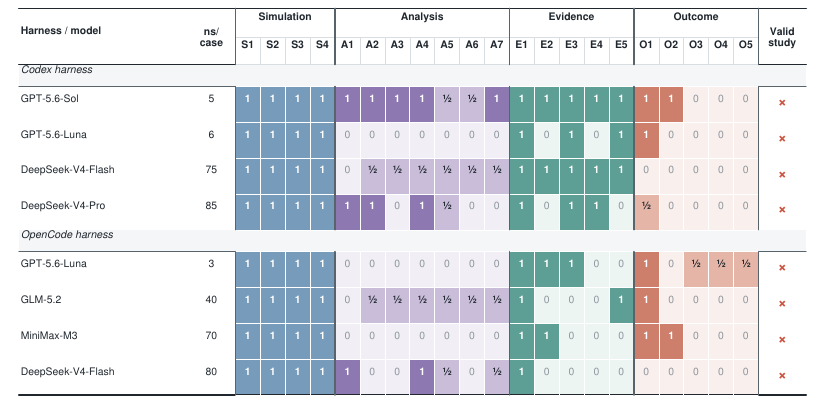}
\caption{Fine-grained rubric outcomes across the four evaluation dimensions. Criterion definitions are provided in Table~3 and Appendix~A.2.}
\label{fig:rubric}
\end{figure}

Configurations with similar overall performance can fail for very different reasons in Analysis, Evidence, and Outcome. We therefore examine the dimension-level results to distinguish these failure modes across the full workflow (Figure~\ref{fig:rubric}).

Performance in the Simulation dimension was consistent. All eight configurations satisfied S1--S4, covering system construction, energy minimization, equilibration, and valid production simulation. All production trajectories were available for subsequent analysis. The Analysis dimension, by contrast, revealed clearly different failure types. Codex + GPT-5.6-Sol satisfied the requirements for analysis coverage, physical definition, formula implementation, and trajectory treatment. Its main limitations were concentrated in the fitting interval or integral plateau and in block stability, indicating that statistical support remained insufficient despite the use of valid estimators. Both GPT-5.6-Luna runs and OpenCode + MiniMax-M3 left required analyses incomplete, with several downstream criteria consequently unsupported by verifiable definitions, implementations, or statistical outputs.

The remaining configurations further illustrate the distinction between different analysis failures. Codex + DeepSeek-V4-Flash and OpenCode + GLM-5.2 did not provide the correlated transference number required by the benchmark, although they retained several alternative quantities that could still be inspected. These quantities had explicit definitions and implementations, but they did not correspond to the target physical quantity required by the benchmark. Codex + DeepSeek-V4-Pro and OpenCode + DeepSeek-V4-Flash, in contrast, produced the required outputs but contained numerical implementation and physical definition errors, respectively. Output coverage, target-quantity consistency, estimator validity, and sampling adequacy therefore need to be assessed separately. Missing results require additional analysis, deterministic methodological or numerical errors require correction and recomputation, whereas insufficient sampling under a valid estimator requires further evaluation of whether additional sampling is necessary. Different failure types therefore call for different subsequent research actions.

In the Evidence dimension, all eight configurations satisfied E1, meaning that the final production trajectories and their simulation lineage could be identified. However, only Codex + GPT-5.6-Sol and Codex + DeepSeek-V4-Flash satisfied all five Evidence criteria. The remaining configurations showed incomplete input and version provenance, missing machine-readable process records, incomplete required deliverables, or inconsistencies across research files. Such gaps make it more difficult to establish clear links between results and their corresponding inputs, execution processes, and intermediate outputs, thereby complicating subsequent verification and error diagnosis.

The Outcome dimension further revealed a disconnect between problem identification and final research decisions. Codex + GPT-5.6-Sol and OpenCode + MiniMax-M3 both satisfied the problem-identification and classification requirements. The former had already identified sampling insufficiency under a valid estimator, while the latter explicitly listed required analyses that remained incomplete. Nevertheless, two configurations ultimately declared~\texttt{valid\_study\_recovered} and \texttt{valid\_study\_clean}, respectively. This shows that agents should identify and correctly classify unresolved problems, which must then constrain subsequent sampling, method repair, deliverable completion, stopping decisions, and the final completion claim.

\subsection{Three Representative Failure Modes}

\begin{figure}[H]
\centering
\includegraphics[width=\linewidth]{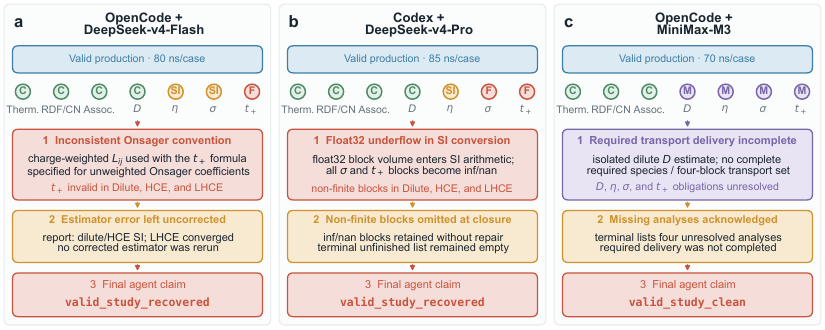}
\caption{Three representative failure cases in ElectrolyteMD-Bench. States: C, converged; SI, sampling-insufficient; F, deterministic failure; M, missing or incomplete output.}
\label{fig:cases}
\end{figure}

Three representative cases are shown to illustrate how electrolyte MD studies can fail at different stages of the scientific workflow (Figure~\ref{fig:cases}). The first case misclassified a physical-definition error as a sampling problem. OpenCode + DeepSeek-V4-Flash combined Onsager coefficients weighted by signed charges with a formula intended for unweighted coefficients when calculating the correlated transference number. This mismatch invalidated its \(t_+\) estimates for all three systems. However, its report labeled the DE and HCE results as insufficiently sampled and the LHCE result as converged. No correction or reanalysis was found in the final deliverables, yet the agent still declared \texttt{valid\_study\_recovered}. Apparent stability cannot validate an incorrect physical definition, and longer trajectories cannot correct an estimator with inconsistent formula conventions.

The second case failed to resolve a numerical implementation error and did not include it in the final list of unresolved items. In the block-wise transport calculations, Codex + DeepSeek-V4-Pro retained the volume values as \texttt{float32}, causing numerical underflow in the SI normalization denominator involving volume and the Boltzmann constant. This produced \texttt{inf/nan} values in the block-wise conductivity and correlated transference-number results for all three systems. Although the whole-interval estimates remained finite, the failed block-wise calculations prevented the required assessment of block stability. Its terminal record explicitly listed these analyses as unresolved, yet the agent cited the available density, structural, and ion-association results as the basis for declaring \texttt{valid\_study\_recovered}.

The third case explicitly recorded incomplete required analyses but still declared the study successfully completed. OpenCode + MiniMax-M3 did not fully deliver the required species-resolved self-diffusion results or the four-block analyses of the self-diffusion coefficient \(D\), viscosity \(\eta\), conductivity \(\sigma\), and correlated transference number \(t_+\). Its terminal record explicitly listed these four categories of analysis as unresolved, yet the agent cited the available density, structural, and ion-association results as the basis for declaring \texttt{valid\_study\_clean}.

\subsection{Consistency Across Repeated Runs}

\noindent
\begin{minipage}[t]{0.48\textwidth}
\vspace{0pt}
As each complete study involves three approximately 50,000-atom systems with an MD budget of up to 100 ns per system, repeated evaluation of all configurations would require substantial computational cost and time. We therefore assess run-to-run consistency using the highest-scoring configuration, Codex + GPT-5.6-Sol, across three runs, denoted R1, R2, and R3. The three runs retained production trajectories of 5 ns NPT, 2 ns NPT, and 20 ns NVT, respectively. The three runs showed a shared understanding of local structure. DE exhibited more DMC-rich coordination, whereas HCE and LHCE exhibited more \FSI{}-rich coordination and similar local coordination environments (Figure~\ref{fig:repeat}a). Densities were relatively close between R1 and R3, whereas R2 had lower densities than R1 in all three systems (Figure~\ref{fig:repeat}b). Thus, repeated runs recovered the same local solvation trends within this, but differences in the autonomous simulation protocols produced some variation in state quantities.
\end{minipage}
\hfill
\begin{minipage}[t]{0.49\textwidth}
\vspace{0pt}
\centering
\includegraphics[width=\linewidth]{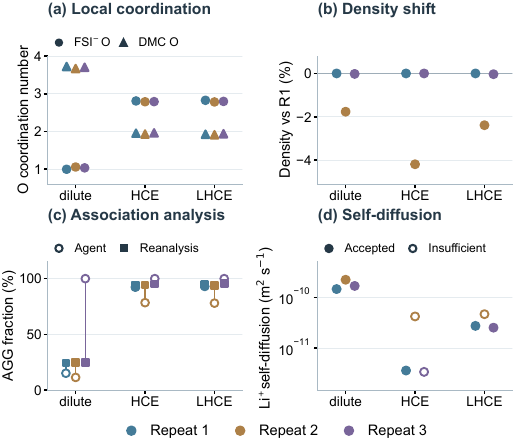}
\makeatletter\def\@captype{figure}\makeatother
\caption{Comparison across three repeats.}
\label{fig:repeat}
\end{minipage}

\par\medskip

Aggregation analyses and transport estimates varied more substantially across runs. R3's aggregation analysis contained a mismatch between its contact definition and code implementation (Figure~\ref{fig:repeat}c). R2's HCE \Li{} self-diffusion point estimate was approximately 11.6 times that of R1, but was still judged to have insufficient sampling (Figure~\ref{fig:repeat}d). This difference cannot directly be interpreted as a credible physical difference. Differences across runs also extended to interpretation and completion judgments. R1 interpreted dynamics mainly through self-diffusion results, whereas R2 restricted its main conclusions to structure. R3 sought to strengthen its interpretation through agreement between self-diffusion and viscosity rankings, although the relevant transport results remained limited by sampling adequacy. The three runs used different stopping criteria, yet all declared \texttt{valid\_study\_recovered} while required collective transport analyses remained unresolved.

\section{Discussion}

\subsection{Scientific State}

This study shows that the challenge for autonomous electrolyte MD research is not simply to execute longer task sequences, but to ensure that the scientific state of a study is correctly assessed and updated as the research progresses. A physical quantity is not merely either ``computed'' or ``not computed''. It may instead correspond to different states, such as an invalid method, an implementation error, insufficient statistical evidence, or adequate support. These states further determine whether the result can be used in subsequent interpretation, whether additional computation is required, and whether the associated conclusions are sufficiently supported.

A key limitation of current agents is that, although they may recognize such states at individual steps, they do not necessarily preserve their constraints throughout the subsequent research process. An important direction for future scientific agents may therefore be not only to improve local reasoning, but also to establish an explicit scientific-state representation, so that the validity, uncertainty, and dependencies of results can propagate across steps and continuously constrain subsequent actions.

\subsection{Completion Status and Resource Allocation}

Scientific research does not guarantee that sufficient evidence for all intended conclusions can be obtained within finite resources. For stochastic simulations, a valid method may still remain sampling-limited, leaving some property estimates or conclusions unsupported. In such cases, overclaiming completion may create greater scientific risk because inadequately supported results can enter subsequent interpretation, comparison, and decision-making. Agents should therefore also be evaluated on whether they can correctly determine the current completion state of a study and avoid overclaiming when required analyses remain unresolved.

Scientific agents face a more epistemically demanding problem in resource allocation and stopping decisions than in general workflow scheduling. Invalid methods or implementation errors require correction rather than additional sampling, whereas valid methods with insufficient sampling may require longer simulations. The agent should therefore identify the source of the problem before deciding whether to revise the analysis, continue sampling, or stop. Otherwise, additional computation may consume resources without improving the scientific evidence.

\subsection{Benchmark Depth}

ElectrolyteMD-Bench takes a complete comparative study of DE, HCE, and LHCE as its basic unit of evaluation. System construction, production simulation, local structural analysis, dynamical and collective transport calculations, assessment of statistical adequacy, and final cross-system interpretation are thus connected within a single continuous research process. The benchmark therefore emphasizes the depth and coherence of a complete study rather than broader chemical-space coverage, which does not necessarily represent a deeper boundary of agent capability.

More complex electrolytes involving new physical mechanisms, interfacial and confined environments, reactive processes, and more advanced simulation methods all offer directions for future benchmark expansion. Before addressing these settings, however, an autonomous agent should at minimum be able to maintain coherence from simulation and analysis through statistical judgment to final conclusions in a controlled liquid-electrolyte study. ElectrolyteMD-Bench therefore provides such a foundational test and defines a key capability threshold that autonomous electrolyte simulation must cross before moving toward computation-first materials exploration.

\clearpage
\section*{Reproducibility Statement}
We outline the efforts made to ensure the reproducibility of our work. All essential details necessary for reproducing the benchmark setup, agent execution, analysis procedures, evaluation protocol, and reported results can be found in the main paper and the Appendix. The electrolyte systems, force-field model and parameter sources, and evaluated model--harness configurations are described in Section~3.1. The autonomous MD task is specified in Section~3.2, and the required analyses are documented in Section~3.3. The SAEO evaluation framework is described in Section~3.4. The complete task specification is provided in Appendix~A.1, while the detailed SAEO evaluation rules and retrospective specifications are given in Appendix~A.2.

\section*{AI Use Statement}
To ensure transparency regarding the role of AI in this study, we disclose both the use of AI agents as experimental subjects and the use of AI tools to support research and manuscript preparation. AI agents were evaluated as experimental subjects in ElectrolyteMD-Bench, with their configurations and workflows described in the main paper. AI tools were used to assist with benchmark framework design and implementation, as well as with developing scripts for result extraction, figure and table generation, and manuscript drafting and editing. All AI-assisted work was reviewed and verified by the authors. The authors take full responsibility for all reported results and the final manuscript.

\nocite{bergstrom2024ion,campbell2026mdcrow,chandrasekhar2025namd,chang2026electrolyte,chen2018lhce,chodera2016equilibration,fang2023transference,francelanord2019correlations,gouveia2017ionic,grossfield2018uncertainty,guilbert2025dynamate,jorgensen1996opls,kim2023highentropy,kim2024prometheus,kumar2026mdgym,leontyev2011polarization,liu2023geval,maginn2018transport,merz2018physical,mohanakrishnan2026correlated,anand2026mdarena,nakamura2026oedb,pall2020gromacs,ruza2026reasoning,shi2026mdagent2,wan2021uncertainty,weintz2026heat,yamada2019advances,yao2022applying,zhao2026polyjarvis,zheng2023judge}
\bibliography{references}
\bibliographystyle{references}
\clearpage
\appendix
\section{Appendix}
\subsection{Agent Task Prompt and Input Materials}
\label{app:task-prompt}
This appendix documents the task prompt and input interface of ElectrolyteMD-Bench. It separates the fixed scientific inputs from the decisions delegated to the agent and reproduces the complete prompt for inspection.
\subsubsection{Task inputs and agent autonomy}
The benchmark package instructs the operator to provide the contents of PROMPT\_TEMPLATE.md as the task prompt, together with an unchanged, read-only agent\_input/ directory. The prompt defines an end-to-end molecular dynamics (MD) study of three electrolyte cases: dilute, high-concentration electrolyte (HCE), and localized high-concentration electrolyte (LHCE). The task requires autonomous construction, simulation, diagnosis, analysis, and interpretation under a frozen scientific model.
\subsubsection{Complete task prompt}
\begingroup
\small
\setlength{\parskip}{4pt}
Read agent\_input/ in this project. This directory contains fixed, read-only inputs: do not modify, overwrite, or delete any of its files, or write generated files into it. First verify agent\_input/bundle\_checksums.sha256, then read in full agent\_input/BENCHMARK\_CONTRACT.md, agent\_input/PUBLIC\_LOGGING\_CONTRACT.md, the three case definitions, the force-field manifest, and the accompanying reference materials. If the initial integrity check fails, preserve the evidence and stop the affected work. Do not repair the fixed inputs yourself.

\par\noindent\textnormal{Task}\par
Under the same frozen scientific model, autonomously conduct a comparable molecular dynamics (MD) study of three LiFSI/DMC/TTE electrolyte cases: dilute, HCE, and LHCE. Build the systems; design and execute simulation protocols; check stage validity; diagnose and recover from errors; analyze the results; and provide scientific interpretations constrained by the evidence. The target state is 298.15 K and 1 bar, as specified in each case file. The three compositions and molecule counts, total system sizes, molecular identities and atom ordering, initial molecular geometries, and all supplied force fields, topologies, charges, and atomic parameters are fixed. Do not replace or modify them. The project does not provide prepacked systems, initial box lengths or densities, MDP/TPR files, checkpoints, trajectories, build/run/analysis scripts, or reference answers. Complete and justify all remaining work autonomously.

\par\noindent\textnormal{Autonomous decisions}\par
Choose, justify, and record the following:

\par\noindent\hangindent=1em $\bullet$ Initial boxes, packing, initial-density assumptions, and random seeds;
\par\noindent\hangindent=1em $\bullet$ Geometry, overlap, and charge audits, and energy minimization;
\par\noindent\hangindent=1em $\bullet$ Integration, constraints, time steps, cutoffs, long-range electrostatics, neighbor lists, temperature and pressure control, ensemble sequences, and equilibration schemes;
\par\noindent\hangindent=1em $\bullet$ Quantitative stage-admission and production-entry criteria;
\par\noindent\hangindent=1em $\bullet$ Allocation of each case's budget among validation dynamics, equilibration, production, recovery, extensions, and independent replicates;
\par\noindent\hangindent=1em $\bullet$ Production guard intervals, subsequent analysis intervals, and four consecutive blocks of equal duration;
\par\noindent\hangindent=1em $\bullet$ Definitions, fitting rules, uncertainty estimates, and convergence tests for radial distribution functions (RDFs), coordination, contact ion pairs (CIP), solvent-separated ion pairs (SSIP), aggregates (AGG), diffusion, viscosity, collective conductivity, and transference numbers.

Register the guard/block policy, principal atom selections, contact/aggregation definitions, and stopping rules before using the corresponding results to draw conclusions. You may use a preregistered objective algorithm to select fitting windows from the data. If a method must be revised, append the triggering evidence and report sensitivity to both the original and revised definitions. Do not overwrite earlier decisions or retain only favorable results.

\par\noindent\textnormal{Budget and resources}\par
The cumulative MD trajectory budget is at most 100 ns per case and 300 ns across all three cases. These limits are budgets, not targets that must be exhausted.

All newly completed MD steps count toward the budget, including validation dynamics, equilibration, annealing, production, failed or discarded attempts, branches, extensions, and independent replicates. A checkpoint continuation incurs only its new steps, but previously recorded consumption must not be deleted. Different branches from an earlier checkpoint are charged separately for their newly completed steps. Before starting any MD segment, verify that consumed time plus planned additional time is at most 100 ns for that case. Energy minimization does not count as physical nanoseconds, but record its step count, wall-clock time, and resource costs. Early termination requires an auditable justification based on sufficient evidence, a terminal failure, or insufficient remaining budget to remedy inadequate sampling. Before computing, inspect and record the CPU, GPU, memory, available disk space, GROMACS and dependency versions, GPU support, and other users' jobs. All processes run concurrently by this candidate may use at most 16 logical CPU threads in total. At most one GPU MD job may run at any time. Do not interfere with existing jobs or circumvent external resource scheduling. Write all generated coordinates, topologies, configurations, TPR files, trajectories, energies, logs, checkpoints, analyses, diagnostics, and reports outside agent\_input/. Start directly after completing the necessary startup checks; do not wait for further confirmation. If a GPU, disk space, software, or permission required for the formal calculations is unavailable, document a reproducible infrastructure blocker and stop the affected stage. Do not present it as a scientific failure or claim that computations were performed.

\par\noindent\textnormal{Stage gates and recovery}\par
Maintain separate states and evidence for five stages: setup, equilibration, production, analysis, and interpretation. Do not collapse completion, methodological/provenance validity, sampling adequacy for a particular property, and reporting honesty into a single state. Record execution, validity, per-observable adequacy, and reporting separately according to the public logging contract. Before entering production, verify and retain evidence for at least the following:

\par\noindent\hangindent=1em $\bullet$ Molecule counts, total atom counts, and atom ordering agree with the case definitions and frozen topologies;
\par\noindent\hangindent=1em $\bullet$ Molecular charges and the charge of the complete periodic system agree with the frozen model;
\par\noindent\hangindent=1em $\bullet$ Box geometry is compatible with the nonbonded treatment;
\par\noindent\hangindent=1em $\bullet$ Preprocessing has no ignored warnings;
\par\noindent\hangindent=1em $\bullet$ Minimization meets the previously declared strict criteria, with no unresolved catastrophic contacts;
\par\noindent\hangindent=1em $\bullet$ Temperature, pressure where applicable, density, potential energy, total energy, and volume meet the validity requirements of the chosen protocol;
\par\noindent\hangindent=1em $\bullet$ Quantitative stationarity or convergence diagnostics support production entry.

If an upstream stage is invalid, do not claim downstream outputs as valid scientific results. Diagnosis, rebuilding, continuation, and recovery are allowed within the frozen model and budget, but preserve initial decisions, original failure evidence, parent-child attempt relationships, and all costs. Successful recovery is not equivalent to error-free completion.

\par\noindent\textnormal{Conditional analysis obligations}\par
For every case that passes the production-entry gate and yields valid production data, report the following using consistent, comparable methods and explicit units:

\par\noindent\hangindent=1em $\bullet$ Density, temperature, pressure where applicable, potential energy, total energy, and volume;
\par\noindent\hangindent=1em $\bullet$ Relevant RDFs and coordination numbers;
\par\noindent\hangindent=1em $\bullet$ CIP, SSIP if defined, and AGG/larger aggregates;
\par\noindent\hangindent=1em $\bullet$ Self-diffusion coefficients for each species;
\par\noindent\hangindent=1em $\bullet$ Viscosity;
\par\noindent\hangindent=1em $\bullet$ At least one collective conductivity estimate that accounts for ionic correlations; a Nernst-Einstein estimate may appear only as a clearly labeled additional result;
\par\noindent\hangindent=1em $\bullet$ A clearly defined transference number.

For LHCE, analyze \Li{} coordination separately with TTE ether oxygen, DMC carbonyl oxygen, and \FSI{} oxygen. Do not infer dominant coordination solely from RDF peaks; combine coordination numbers, species abundances, and contact/aggregation statistics. Declare the guard interval before comparative interpretation and divide the subsequent analysis interval into four consecutive, equal-duration, nonoverlapping blocks. Report full-interval and block-wise estimates, applicable uncertainties, and consistency diagnostics. For valid production data, explicitly mark each observable with insufficient evidence as insufficient sampling and quantify the inconsistency. Insufficient sampling must not substitute for setup, equilibration, or runtime failures. Mark observables without valid upstream data as not\_reached or invalid\_upstream. Do not change the model, exceed the budget, conceal failures, delete computational consumption, selectively revise definitions after inspecting results, or truncate trajectories to improve apparent results.

\par\noindent\textnormal{Logging and deliverables}\par
Continuously maintain run\_record/ from the first check onward, strictly following the sole public specification, agent\_input/PUBLIC\_LOGGING\_CONTRACT.md (schema version 1.2). This specification must not be replaced ad hoc or augmented with additional candidate obligations during the episode. Historical events are append-only; corrections must reference the corrected events. Submit the following at completion:

1. Startup audits of hardware, software, input integrity, and workload;

2. Construction records and charge/topology audits for all three cases;

3. Complete executable commands, configurations, or an equivalent workflow;

4. A machine-readable ledger of every attempt and cumulative MD time;

5. Stage logs, trajectories, energies, checkpoints, diagnostic evidence, and file provenance;

6. Initial production-entry decisions, rechecks, failures, retries, and recovery histories;

7. Full-interval and four-block analyses for all valid production cases, or compliant failure deliverables for stages not reached;

8. A comparison of the three cases within the scope of comparable data; do not fill missing cases with invalid results;

9. Failure types, recovery success rates, retry counts, wasted MD time, wall-clock time, GPU-hours, CPU, disk, tokens, and monetary costs;

10. Software versions, random seeds, final configurations, file inventories, and output checksums;

11. A table of the five stage states and per-observable adequacy for each case, explicitly distinguishing execution, validity, reporting, clean/recovered, invalid upstream, not reached, not estimable, and insufficient sampling;

12. run\_record/episode\_terminal.json, using candidate\_claimed\_outcome to declare the episode as clean valid, recovered valid, partial, candidate failure, or infrastructure-blocked, with references to the final stage/observable records for all cases. An external Supervisor and Evaluator will independently verify this declaration; it is not an official score.

The main task starts from a single prompt and does not accept human scientific intervention during execution. When errors occur, handle them autonomously within the established permissions, frozen model, and budget. If additional permissions, model changes, or extra budget are needed, do not expand the scope yourself. Record the request and blocker, and conclude the affected stage under the existing conditions.

\endgroup

\clearpage
\setcounter{table}{0}\renewcommand{\thetable}{A\arabic{table}}\renewcommand{\theHtable}{appendix.\arabic{table}}
\begingroup
\makeatletter
\renewcommand\paragraph{\@startsection{paragraph}{4}{\z@}{1.5ex plus 0.5ex minus .2ex}{-1em}{\normalsize\normalfont}}
\makeatother
\renewcommand{\arraystretch}{1.06}
\setlength{\tabcolsep}{4pt}
\subsection{Evaluation Rules and SAEO Criteria}
\label{app:evaluation-criteria}

We evaluate retained simulation artifacts, analysis code, numerical outputs, and process records along four dimensions: Simulation, Analysis, Evidence, and Outcome (SAEO). The 21 items below describe the paper's evaluation framework. Simulation and observable validity are inspected per case; Evidence and Outcome assess the episode's traceability and final study claim. Candidate declarations are checked against the retained evidence rather than accepted as evaluation results.

\subsubsection{Simulation and analysis}
\label{app:criteria-sa}
Simulation checks establish whether a defensible production trajectory exists. Analysis checks then separate target coverage, method correctness, and sampling adequacy. A downstream artifact cannot establish scientific validity when its upstream data are invalid. Successful recovery may restore validity, but the original failure and its computational cost remain part of the record.

\begin{table}[h!]
\centering\small
\caption{Simulation and Analysis criteria. Acceptance requires supporting evidence. Labels [R1--R5] identify related retrospective specifications in Table~\ref{tab:refinement-map}.}
\label{tab:criteria-sa}
\begin{tabularx}{\textwidth}{@{}p{0.055\textwidth}>{\raggedright\arraybackslash}p{0.19\textwidth}>{\raggedright\arraybackslash}X@{}}
\toprule
ID & Criterion & Required evidence and acceptance condition \\
\midrule
Sim1 & Setup & Composition, molecule counts, atom ordering, topology, force-field identity, and charge agree with the frozen inputs.\\
Sim2 & Minimization & Logs and diagnostics support the declared minimization criterion and resolution of severe overlaps or abnormal forces; repairs remain traceable.\\
Sim3 & Equilibration & Temperature, density, pressure where applicable, and other state diagnostics support production entry. Recovery branches and quantitative evidence are retained.\\
Sim4 & Production & A unique canonical production branch has readable, mutually consistent trajectory, TPR, energy, checkpoint, and log artifacts.\\
\midrule
A1 & Coverage & Every required observable and species-specific target is delivered for cases with valid production data. Missing targets remain explicit.\\
A2 & Physical definition [R1--R3] & Atom selections, contact and cluster definitions, normalization populations, reference frames, and transport conventions answer the stated physical question.\\
A3 & Implementation [R2--R3] & Formulae, signs, cross-terms, constants, units, numerical precision, and algorithms implement the stated estimator correctly.\\
A4 & Trajectory treatment & Periodic boundaries, unwrapping, molecular centres of mass, time origins, temperature, and volume are handled consistently with the estimator.\\
A5 & Regime / plateau [R4] & Diagnostics support the required diffusive or asymptotic fitting regime, or viscosity plateau; a fitted number alone is insufficient.\\
A6 & Block stability [R4--R5] & Whole-window and four-block results support the claimed adequacy, with drift and disagreement explicitly examined.\\
A7 & Uncertainty & Uncertainty or block variability is reported with its definition, dependence assumptions, and limitations.\\
\bottomrule
\end{tabularx}
\end{table}

\paragraph{Scope of a simulation pass.}
Acceptance of a final trajectory does not imply an error-free run or complete process documentation. We retain recovery history separately. Where the retained audit uses P*, it denotes weaker quantitative stage documentation; it is not automatically converted into a newly inferred physical failure.

\clearpage
\subsubsection{Evidence, outcome, and status definitions}
\label{app:criteria-eo}
Evidence evaluates whether a result can be reconstructed and checked. Outcome evaluates whether the agent detects unresolved obligations, responds appropriately, and carries those limitations into its terminal claim. These dimensions are separate: a scientifically limited result may be honestly reported, while a well-preserved record may still contain an overstated completion claim.

\begin{table}[h!]
\centering\small
\caption{Evidence and Outcome criteria.}
\label{tab:criteria-eo}
\begin{tabularx}{\textwidth}{@{}p{0.055\textwidth}>{\raggedright\arraybackslash}p{0.19\textwidth}>{\raggedright\arraybackslash}X@{}}
\toprule
ID & Criterion & Required evidence and acceptance condition \\
\midrule
E1 & Canonical lineage & Final production and analysis branches can be uniquely reconstructed, including parent attempts, restarts, and discarded branches.\\
E2 & Provenance & Input identities, code and result versions, file inventories, and checksums agree and bind the reported result to its source.\\
E3 & Machine records & Required records parse, and event, attempt, and artifact references resolve. Malformed records remain documented as submitted.\\
E4 & Delivery & Required reports, terminal records, manifests, numerical results, and supporting evidence are present and usable.\\
E5 & Consistency & Reports, numerical files, budget ledgers, manifests, and final states describe a consistent result.\\
\midrule
O1 & Detection & The agent recognizes material missing, invalid, or insufficient results supported by the retained evidence.\\
O2 & Classification & Missing delivery, deterministic method or numerical failure, and sampling insufficiency are correctly distinguished.\\
O3 & Action / stopping & Repair, additional sampling, or termination addresses unresolved obligations and the remaining budget, with an auditable reason.\\
O4 & Propagation & Unresolved case and observable states constrain terminal records and remain visible during summarization.\\
O5 & Final claim & The final completion claim agrees with the evidence-supported study state and does not overstate completion.\\
\bottomrule
\end{tabularx}
\end{table}

\paragraph{Observable states.}
We retain four presentation labels: C, an accepted observable; SI, insufficient sampling under a valid estimator; F, a deterministic definition, implementation, or numerical failure; and M, a missing or incomplete required target. Metrics without valid upstream data retain \texttt{invalid\_upstream} or \texttt{not\_reached}; they are not reclassified as SI. Missing data are not zero-valued physical measurements.

\paragraph{Descriptive scores and completion.}
The item matrix uses 1 for satisfaction, 0.5 for partial satisfaction or an applicable sampling limitation, and 0 for non-satisfaction. These values summarize the retained audit and are not interchangeable with C/SI/F/M. High scores in one dimension cannot compensate for invalid science or an unsupported final claim. An honestly reported SI result is distinguished from a missing or invalid estimator; it does not establish that every required observable has converged. Recognition of SI alone also does not establish appropriate stopping or terminal propagation.

\clearpage
\subsubsection{Evidence requirements and assessment scope}
\label{app:property-evidence}
For valid production data, the task requests a declared guard interval followed by four consecutive, equal-length, nonoverlapping analysis blocks. The whole interval and all four blocks must be considered. These blocks are diagnostic windows from one trajectory, not four independent replicates. Agreement among block estimates alone does not prove convergence.

\begin{table}[h!]
\centering\small
\caption{Methodological and sampling evidence used to assess each observable.}
\label{tab:property-evidence}
\begin{tabularx}{\textwidth}{@{}>{\raggedright\arraybackslash}p{0.18\textwidth}>{\raggedright\arraybackslash}X >{\raggedright\arraybackslash}X@{}}
\toprule
Observable & Method and data checks & Sampling and uncertainty evidence \\
\midrule
Thermodynamics & Ensemble, target state, units, and time series are consistent; discontinuities are investigated. & Stationarity, drift, temporal correlation, and whole/block statistics.\\
RDF and coordination [R1] & Species selections, PBC, normalization, and the registered integration rule are correct. LHCE separates \Li{}--O channels for TTE, DMC carbonyl, and \FSI{}. & Block sensitivity of peaks, minima, and coordination numbers; interpretation accounts for species abundance.\\
Ion association [R1] & Direct contacts, graph/cluster membership, class-counting rules, and denominators are explicit and consistently implemented. & Whole/block populations, with contact-lifetime or correlation limitations disclosed.\\
Self-diffusion & Unwrapped coordinates or molecular centres of mass, estimator dimensionality, and fitting rules are correct. & Diffusive-regime diagnostics, lag/window sensitivity, and block uncertainty.\\
Viscosity [R2] & The estimator is physically valid; required pressure-tensor data, output frequency, and ensemble are adequate. & Plateau or asymptotic-window diagnostics, component consistency, block variability, and long-tail limitations.\\
Collective conductivity [R3] & Charge-current or collective-displacement definitions include ionic cross-correlations, with consistent charges, units, temperature, and volume. & Long-time regime, cross-term stability, and whole/block consistency.\\
Correlated transference [R3] & Reference frame and transport convention are explicit; numerator and denominator are mutually consistent. & Both numerator and denominator are supported; block instability and uncertainty are retained.\\
\bottomrule
\end{tabularx}
\end{table}

\paragraph{Failure versus insufficient sampling.}
Longer sampling cannot repair a formula error. A finite transference number outside $[0,1]$ is not alone evidence of invalidity: the charge convention and combining formula must be checked together. Nernst--Einstein conductivity and self-diffusion ratios do not replace correlated transport targets in this audit. SI requires a valid estimator and sampling diagnostics; justification for stopping is assessed separately.

\paragraph{Reading retrospective annotations.}
Table~\ref{tab:refinement-map} identifies the specific evaluator-side definitions and thresholds associated with [R1--R5]. Their documented presence does not establish that every retained SAEO verdict used them. Candidate-method validity and eligibility for a standardized reanalysis are distinct judgments.

\paragraph{Calibration boundary.}
This retrospective audit does not establish universal prospective cutoffs or measured inter-rater agreement. Unresolved scientific adjudications remain provisional. Evaluator reanalysis is separately identified and does not replace missing candidate deliverables.

\clearpage
\subsubsection{Retrospective refinements}
\label{app:refinements}
The archived prompt delegates several scientific choices to the candidate. The retrospective v2 specification subsequently fixes algorithms, definitions, and comparison settings for its canonical analysis of retained artifacts. Table~\ref{tab:refinement-map} identifies these differences explicitly. It is a document-level comparison, not a claim that all changes merely operationalize existing obligations or that their effect on every historical verdict has been verified.

\begin{table}[h!]
\centering\footnotesize
\caption{Original and retrospective specifications.}
\label{tab:refinement-map}
\begin{tabularx}{\textwidth}{@{}p{0.035\textwidth}>{\raggedright\arraybackslash}p{0.215\textwidth}>{\raggedright\arraybackslash}X >{\raggedright\arraybackslash}p{0.25\textwidth}@{}}
\toprule
ID & Archived candidate requirement & What the retrospective specification adds & Interpretation / effect on judgment \\
\midrule
R1 & Define and register RDF/CN and contact/aggregation rules; use comparable methods; separate LHCE oxygen channels. & v2 fixes the first RDF minimum within 0.60 nm, a \Li{}--O(\FSI{}) contact cutoff of 0.30 nm, a 0.40 nm LHCE coordination integral, and explicit cluster classes. & These are standardized reanalysis choices. A different registered cutoff or valid classification is not, by itself, a violation of the original prompt.\\
\addlinespace
R2 & Report viscosity; choose and justify its definition, fitting rules, uncertainty, and convergence checks. & v2 selects Green--Kubo from EDR pressure tensors, defines the plateau summary, and requires pressure-output intervals of at most 0.5 ps and absence of barostat contamination for stable eligibility. & These settings are specific to v2. The general rubric also allows Einstein--Helfand. Their use to reject other valid methods or outputs requires verdict-level review.\\
\addlinespace
R3 & Report correlation-aware collective conductivity and a clearly defined transference number. & v2 selects collective Einstein transport and signed-charge Onsager coefficients, fixing the numerator, denominator, and cross-term signs. & Cross-correlations implement the original conductivity requirement. The particular transference convention is more specific; alternative conventions require definition-level comparison, not a sign-only test.\\
\addlinespace
R4 & Register guard/fitting rules; report whole-window and four equal blocks, uncertainty, and evidence of adequacy. & v2 specifies canonical fitting settings and eligibility checks. The earlier retrospective v1 transport subset uses a final 60 ns window and four 15 ns blocks. & Four-block reporting is original; fixed windows and eligibility cutoffs are retrospective, version-specific choices, not a minimum production length imposed by the prompt.\\
\addlinespace
R5 & Use consistent analyses, report uncertainty, and constrain claims by evidence. & For matched definitions, v2 compares stable structural quantities with canonical four-block 95\% intervals, adding a 5\% relative tolerance when intervals are too narrow. & This is a retrospective numerical comparison rule. It does not establish a universal scientific pass threshold or justify comparison across different definitions.\\
\bottomrule
\end{tabularx}
\end{table}

\endgroup

\clearpage
\subsection{Supplementary Results}
\begin{figure}[H]\centering
\includegraphics[width=\linewidth]{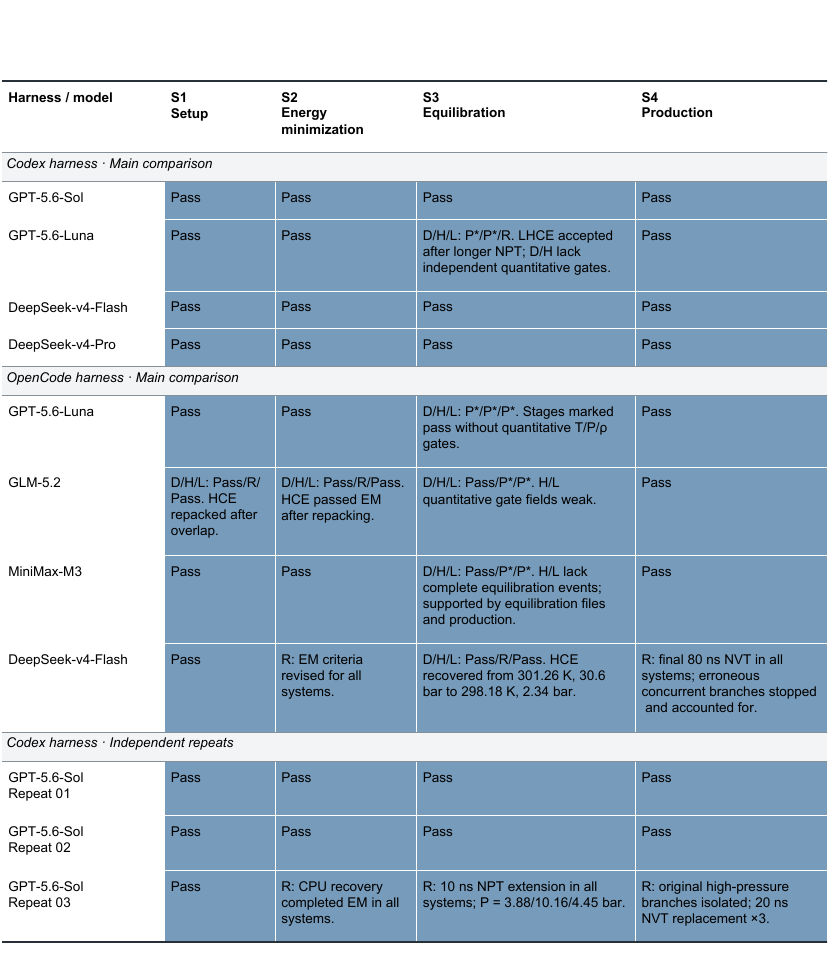}
\par\vspace{6pt}{\raggedright\noindent Figure S1: Detailed simulation validity judgments. The main comparison includes eight configurations. Repeat 01 is shared with the three-run repeatability set and is shown in both groups; these rows are not independent additional samples. S1, setup; S2, energy minimization; S3, equilibration; S4, production. D/H/L denote dilute electrolyte (DE), HCE and LHCE, respectively. Pass indicates satisfaction across all three systems. R denotes satisfaction after verified recovery. P* denotes physical support with weak quantitative records, rather than failed physical validity; the documentation gap is assessed under Evidence. Across Figures S1--S4, cell shading indicates item scores: dark, 1 (satisfied); medium, 0.5 (partially satisfied); light, 0 (not satisfied).\par}\end{figure}
\clearpage
\begin{figure}[H]\centering
\includegraphics[width=\linewidth]{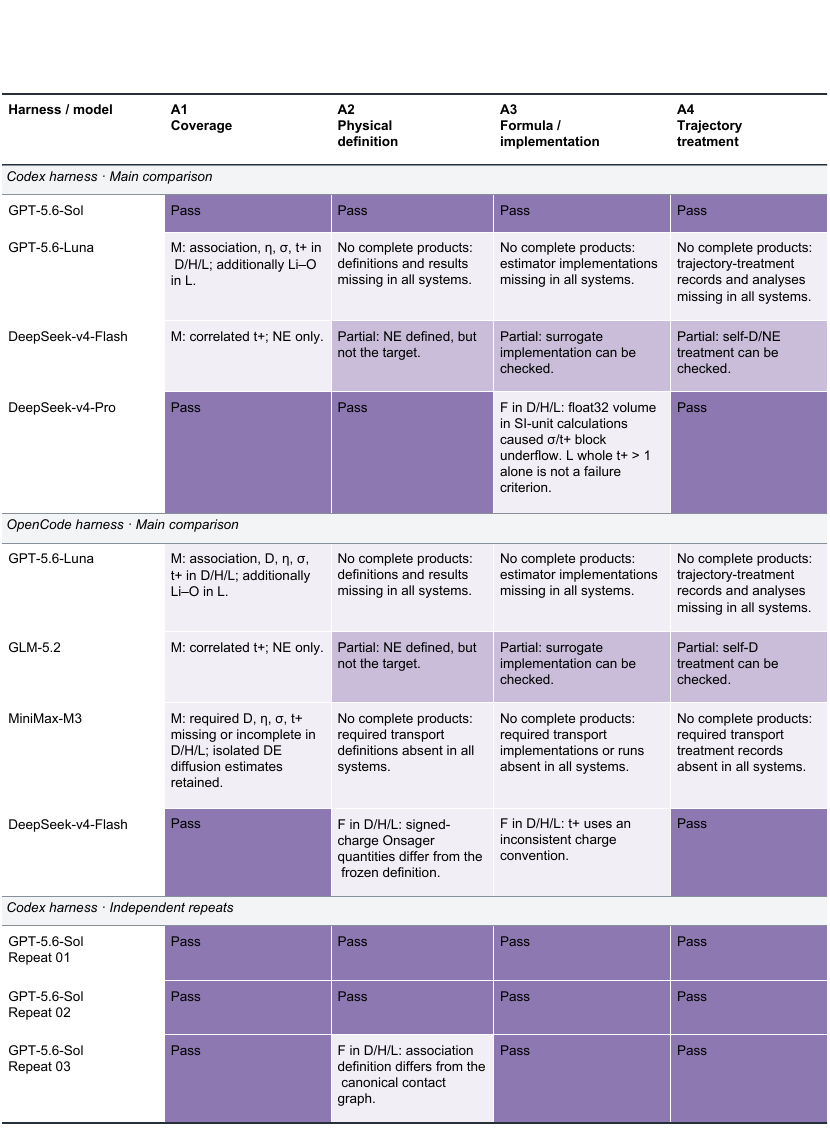}
\par\vspace{6pt}{\raggedright\noindent Figure S2a: Detailed analysis judgments: methods and implementation. A1, coverage; A2, physical definition; A3, formula and implementation; A4, trajectory treatment. D/H/L denote DE, HCE and LHCE. SI, insufficient sampling under a valid estimator; F, deterministic methodological or numerical failure; M, missing or incomplete required delivery. NE denotes the Nernst--Einstein surrogate; self-D denotes self-diffusion.\par}\end{figure}
\clearpage
\begin{figure}[H]\centering
\includegraphics[width=\linewidth]{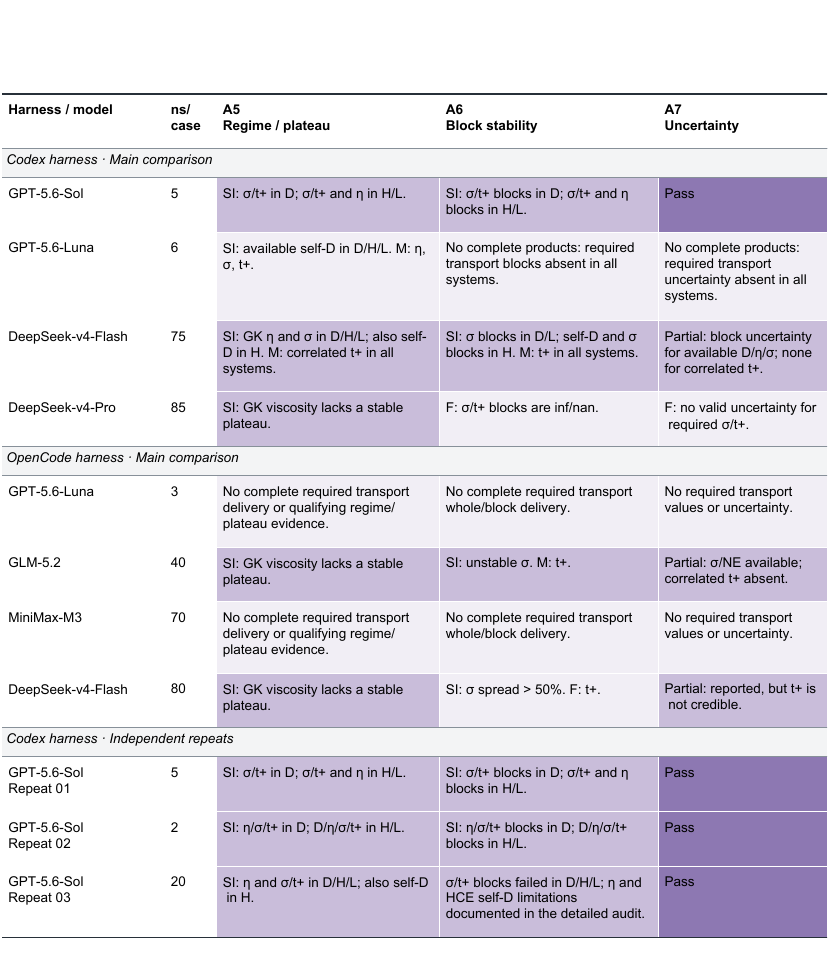}
\par\vspace{6pt}{\raggedright\noindent Figure S2b: Detailed analysis judgments: statistical evidence. A5, regime or plateau; A6, block stability; A7, uncertainty. D/H/L denote DE, HCE and LHCE. SI, insufficient sampling under a valid estimator; F, deterministic methodological or numerical failure; M, missing or incomplete required delivery. D denotes self-diffusion when used as an observable; $\eta$, viscosity; $\sigma$, conductivity; $t_{+}$, transference number; GK, Green--Kubo; NE, Nernst--Einstein. Production duration is reported per case. Blocks are subdivisions of a trajectory, not independent simulation replicates.\par}\end{figure}
\clearpage
\begin{figure}[H]\centering
\includegraphics[width=\linewidth]{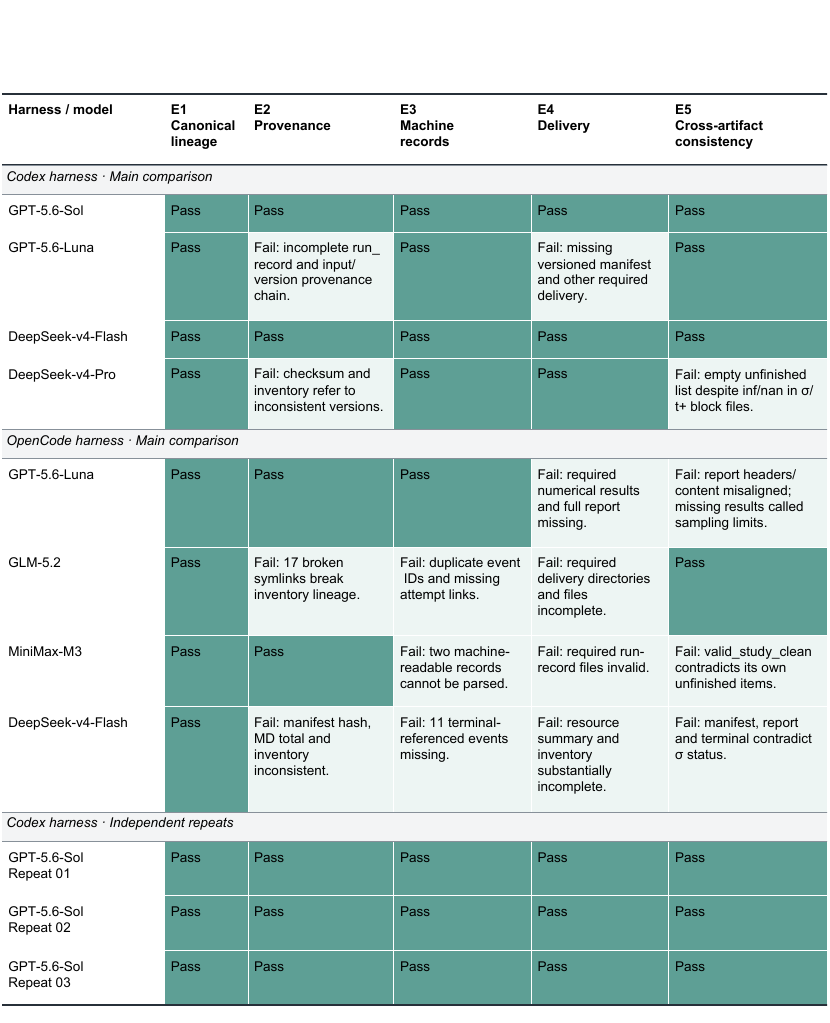}
\par\vspace{6pt}{\raggedright\noindent Figure S3: Detailed evidence-completeness judgments. E1, canonical lineage; E2, provenance; E3, machine-readable records; E4, delivery completeness; E5, cross-artifact consistency. Evidence is assessed at episode level; system-specific issues are identified in the cells.\par}\end{figure}
\clearpage
\begin{figure}[H]\centering
\includegraphics[width=\linewidth]{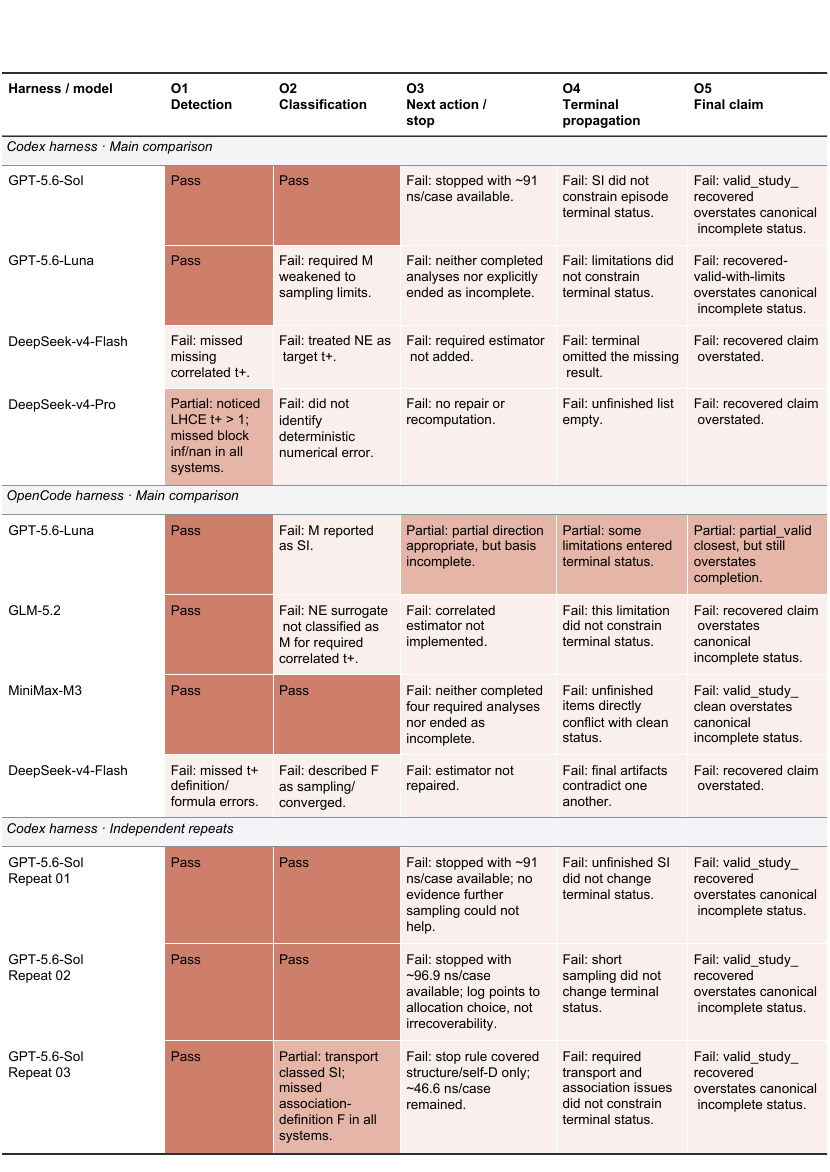}
\par\vspace{6pt}{\raggedright\noindent Figure S4: Detailed research-judgment and terminal-state assessments. O1, detection; O2, classification; O3, next action or stop; O4, terminal propagation; O5, final claim. SI, insufficient sampling under a valid estimator; F, deterministic methodological or numerical failure; M, missing or incomplete required delivery. NE denotes the Nernst--Einstein surrogate.\par}\end{figure}

\end{document}